\documentclass[PRD,showpacs,preprint,floatfix]{revtex4} 
\usepackage{hyperref}
\usepackage{epsfig} 
\usepackage{graphicx}
\usepackage{amsfonts}
\usepackage{amssymb}
\usepackage{amsbsy}
\usepackage{amsmath}
\usepackage{latexsym}

\def\none{\nonumber\\}

\def\be {\begin{equation}}
\def\ee  {\end{equation}}
\def\bea {\begin{eqnarray}}
\def\eea {\end{eqnarray}}

\begin{document}
\noindent
\title{Local Hamiltonian for Spherically Symmetric Collapse: Geometrodynamics Approach}
\author{Jack Gegenberg }
\affiliation{
 Dept. of Mathematics and
Statistics and Dept. of Physics, University of New Brunswick, Fredericton, NB, Canada E3B
5A3\\
AND\\
 Perimeter Institute for Theoretical Physics, 31 Caroline Street North,Waterloo, Ontario, Canada N2L 2Y5}
\author{Gabor Kunstatter }
\affiliation{ Dept. of Physics and Winnipeg Institute for Theoretical Physics, University of Winnipeg, Winnipeg, Manitoba,Canada R3B 2E9\\}
\begin{abstract}
Recently a {\it local} true (completely gauge fixed) Hamiltonian for spherically symmetric collapse was derived in terms of Ashtekar variables. We show that such a local Hamiltonian follows directly from the geometrodynamics of gravity theories that obey a Birkhoff theorem and possess a mass function that is constant on the constraint surface in vacuum. In addition to clarifying the geometrical content, our approach has the advantage that it can be directly applied to a large class of spherically symmetric and 2D gravity theories, including $p$-th order Lovelock gravity in D dimensions.  The resulting expression for the true local Hamiltonian is universal and remarkably simple in form.
\end{abstract}
\date{\today}

\maketitle
\section{Introduction}

Despite considerable progress in recent years in string theory\cite{string} and loop quantum gravity\cite{lqg}, relatively little is known about the quantum dynamics of self-gravitating systems. General covariance provides the classical theory with much of its underlying elegance but it also leads to deep conceptual problems due to the need to fix a gauge (i.e. choose a coordinate system) and the associated problem of time\cite{time}. The situation is further complicated by predictions, primarily from string theory, of quantum induced, higher order curvature terms in the action and the possible existence of more than three spatial dimensions.

Since the groundbreaking work of Berger {\it et. al.}\cite{berger} and Unruh\cite{unruh} in the 1970's spherically symmetric scalar fields  have provided a fruitful testing ground for the classical and quantum dynamics of gravitational collapse. For example, as first shown by Choptuik\cite{choptuik}, spherical collapse exhibits interesting critical behaviour. The Birkhoff theorem, which states in essence that there is no spherically symmetric gravitational radiation, is a key simplifying feature, since there are at most only a finite number of gravitational dynamical modes to consider.  At the quantum level, the Birkhoff theorem guarantees that the vacuum theory can be quantized exactly, and this has been done via many different techniques\cite{vacuum}.

Despite the drastic simplification afforded by the Birkhoff theorem, the dynamics of the matter degrees of freedom is greatly complicated by gravitational self-interactions. The study of the quantum dynamics has to date been rendered intractable by the fact that completely fixing the gauge appears to invariably lead to a non-local Hamiltonian for the matter field\cite{berger, unruh}. (For a current discussion see \cite{viqar2, DGK}.)

A significant step has recently been made  by Alvarez {\it et. al.}\cite{AGP}, who derived a local gauge fixed Hamiltonian to describe the spherically symmetric collapse of a massless scalar field coupled to Einstein gravity in four spacetime dimensions. They described the gravitational field via Ashtekar's connection variables as presented most recently by Bojowald and Swiderski\cite{BS}.

In the following, we prove that geometrodynamical arguments can be used to obtain a similar, but algebraically simpler, local Hamiltonian in a large class of gravity theories that obey a Birkhoff theorem. The suggestion that the analysis is greatly simplified in terms of the standard ADM variables was made independently by V. Husain\cite{viqar} and W. Unruh\cite{bill} and verified explicitly by Unruh for 4D Einstein gravity. We are able to achieve a further simplication by parameterizing the gravitational variables using the mass function and its conjugate. In addition to allowing us to consider simultaneously a large class of theories, our choice of parameterization and gauge yields a reduced Hamiltonian that is very simple in form. This in turn facilitates the discussion of the consistency of the gauge condition for generic initial data.

The theories to which our analysis applies include generic 2D dilaton gravity\cite{grumiller} and $p$-th order Lovelock gravity\cite{lovelock, hideki} in D dimensions\footnote{$p$-th order refers to the highest power of curvature appearing in the action}. Lovelock gravity theories were originally constructed as the most general manifestly covariant higher dimensional versions of Einstein gravity for which the equations of motion are second order in the time coordinate- this ensures reasonable causal behaviour for the propagation of gravitational fields \cite{lovelock}.  There are string theoretic arguments that point to Lovelock gravity  as the low energy limit of consistency conditions for the propogation of strings in curved spacetimes \cite{string lovelock}.  In particular, these stringy consistency conditions contain higher order curvature terms in the same specific linear combinations that occur in Lovelock gravity. If higher dimensional black holes can be probed via the LHC, then it may be possible to rule out certain terms in Lovelock gravity theories \cite{rizzo}. Finally, and most importantly for what follows, Lovelock theories are known to obey a generalized Birkhoff theorem \cite{gen birkhoff, HM11}.

The paper is organized as follows: in the next Section we describe the geometrodynamics and Hamiltonian framework of theories of gravity that obey a Birkhoff theorem and hence possess a mass function\cite{misner_sharpe}. Section III shows in full generality how to gauge fix in such theories so as  to obtain a true local Hamiltonian. Section IV applies the general framework to $p$-th order Lovelock gravity in D dimensions and derives the local Hamiltonian while the following Section discusses boundary conditions. Finally, we close with a summary and prospects for future work.

\section{Geometrodynamics in General}
We will work with the following ADM decomposition using Kuchar's\cite{kuchar} parameters and conventions:
\be
ds^2= -N^2dt^2+\Lambda^2(dr+N_rdt)^2+R^2d\Omega^{D-2},
\label{eq:kuchar adm}
\ee
The metric functions $N,\Lambda, N_r,R$ are arbitrary functions of the two spatial coordinates $(r,t)$, while $R(r,t)$ is the areal radius of the spherically invariant $(D-2)$ spheres. The following analysis also applies directly to gravity in  two space-time dimensions where the role of $R$ is played by a scalar field (the dilaton) non-minimally coupled to the curvature. Standard Hamiltonian techniques reveal that for the spherically symmetric ansatz above any coordinate invariant gravitational Lagrangian can be written, up to boundary terms, in the form:
\be
L^{(G)} = P_\Lambda\dot{\Lambda} + P_R\dot{R} - N H^{(G)}-N_r H_r^{(G)},
\label{eq:action}
\ee
where $H^{(G)}$ and $H^{(G)}_r$ are the gravitational parts of the Hamiltonian and diffeomorphism constraints, respectively. The transformation properties of the ADM variables under changes of spatial coordinates require the gravitational diffeomorphism constraint to be:
\be
H^{(G)}_r = -P_\Lambda'\Lambda + P_R R'.
\ee

Adding a minimally coupled scalar field $\psi$ yields the Lagrangian:
\be
L_{TOT}=L^{(G)}+L^{(m)}=\int dr \left(P_\Lambda\dot{\Lambda} + P_R\dot{R}+\Pi_\psi \dot{\psi} - N (H^{(G)}+\frac{\rho^{(m)}}{\Lambda})-N_r( H_r^{(G)}+H_r^{(m}))\right),
\ee
where
\be
\rho^{(m)}:=\frac{1}{2}\left(\frac{\Pi_\psi^2}{R^{D-2}} +  R^{D-2}(\psi')^2\right),
\label{eq:rho m}
\ee
is the energy density of the scalar field without self-interactions.  This could be generalized to other types of matter subject to the
restriction that it does not contain derivatives of the metric.
The form of the matter part of the diffeomorphism constraint is again dictated by the transformation properties of the fields. For a minimally coupled scalar field $\psi$:
\bea
H^{(m)}_r=\Pi_\psi \psi'.
\eea

In what follows we show that a local Hamiltonian as in \cite{AGP} can be derived using Kuchar' geometrodynamics\cite{kuchar} in a large class of theories that satisfy a Birkhoff theorem, i.e. for which the spherically symmetric vacuum solution is static and is parametrized by a mass function $M$ that is constant on the constraint surface. In a fully dynamical setting any spherically symmetric metric can be written in Schwarzschild-type coordinates as:
\be
d\tau^2 = -F(R;M)N_s^2dT^2 + F^{-1}(R;M) dR^2 + R^2 d\Omega^{D-2}.
\label{eq:schwarz metric}
\ee
where $M=M(R,T)$ is the (generalized) Misner-Sharp mass function \cite{misner_sharpe} and $N_s=N_s(R,T)$ is the lapse function.

In D dimensional spherically symmetric Einstein gravity the Misner-Sharp mass can be written\cite{misner_sharpe}:
\be
M =\frac{R^{D-3}}{2}(1-|\nabla R|^2) = \frac{R^{D-3}}{2}(1-F),
\ee
whereas in general Lovelock gravity it is of the form\cite{hideki}:
\bea
{M}&:=&\sum^{[n/2]}_p \tilde{\alpha}_{(p)}R^{n-1-2p}(1-F)^p.
\label{eq:M1a}
\eea
The mass function $M$ and the areal radius $R$ are invariant under coordinate transformations preserving the spherically symmetric form of the metric.

The essence of geometrodynamics\cite{kuchar} is the construction of the canonical transformation relating the ADM variables $(\Lambda, P_\Lambda),(R,{P}_R)$ to the new phase space variables:
$(P_M,M),(R,\tilde{P}_R)$.  This enables one to extract the physical/geometrical properties of the theory\cite{GKM}. The geometrodynamics of 4D Gauss-Bonnet was first described by Louko {\it et. al.}\cite{LSW} and has recently been completed for $p$-th order Lovelock gravity\cite{TLKM,MKT}. Previous analyses considered only the static, vacuum case, but it also can be shown to go through for the dynamical metric (\ref{eq:schwarz metric}) above\cite{MKT}.

It is useful to define
\be
y:= N^{-1}(\dot{R} - N_r R'),
\label{eq:y}
\ee
 since $\dot{R}$ can only appear in the action in this invariant combination\cite{kuchar}. Doing the coordinate transformation from $(R,T)$ to $(r,t)$ in (\ref{eq:schwarz metric}) and matching coefficients of $dr^2$, etc. in (\ref{eq:kuchar adm}) one obtains the key relations:
\bea
N_s T' &=& \frac{\Lambda y }{F};
\label{eq:PM}\\
F(R;M) &=& \frac{R'^2}{\Lambda^2} - y^2.
\label{eq:FRM}
\eea
As shown originally by Kuchar\cite{kuchar} for vacuum spherically symmetric gravity:
\be
P_M := -\frac{\Lambda y}{F} = - N_s T',
\label{eq:PM2}
\ee
is canonically conjugate to the mass function $M$. The analysis is identical in the dynamical case, since $N_s$ appears only in the relationship between $P_M$ and the Schwarzschild time $T$.

Kuchar\cite{kuchar} argued, based on the form of the diffeomorphism constraints before and after the transformation, that $\tilde{P}_R$ will obey:
\be
\tilde{P}_R   = P_R-P_\Lambda'\Lambda/R' - P_M M'/R'.
\ee
After the change of variables from $(\Lambda,P_\Lambda,R,P_R)$ to $(M,P_M,R,\tilde{P}_R)$, Eq.(\ref{eq:action}) can be written:
\bea
L^{(G)}
 &=& P_M\dot{M} + \tilde{P}_R\dot{R} - N H^{(G)}-N_r H_r^{(G)}.
\label{eq:action2}
\eea

A sufficient condition for Kuchar' geometrodynamics to work is that the gravitational action depend linearly on $\dot{\Lambda}$. This is the case for generic 2D dilaton gravity and for generic Lovelock theory\cite{MKT}. Following  \cite{TLKM} we  write the gravitational lagrangian as :
\be
L^{(G)}= B_1(R,y,\Lambda)\dot{\Lambda}+B_0(R,y,\Lambda).
\label{eq:LG1}
\ee
Clearly
\be
P_\Lambda = B_1(R,y, \Lambda),
\label{eq:PLambda}
\ee
can be inverted to solve for $y$ purely in terms of $\Lambda, P_\Lambda, R$.  That is
\be
y=y(\Lambda,P_\Lambda,R).
\ee
The crucial property of the above is that $y$ does not depend on $P_R$ so that we can use (\ref{eq:PM2}) and (\ref{eq:FRM}) to solve algebraically for $\Lambda$ and $P_\Lambda$ in terms of $M$, $P_M$ and $R$. The Hamiltonian derivable from (\ref{eq:LG1}) then takes the form:
\bea
\cal{H}^{(G)} &=& P_\Lambda\dot{\Lambda} + P_R\dot{R} - L^{(G)}=P_R\dot{R} - B_0(R,y,\Lambda)\none
  &=& N_r P_R R' + N P_R y(\Lambda,P_\Lambda,R)- B_0(\Lambda,P_\Lambda,R),
\label{eq:GHam}
\eea
where we have used the definition of $y$ to get the last line.
Thus the Hamiltonian is linear in $P_R$. The first term will of course contribute to the diffeomorphism constraint while the second will appear in the Hamiltonian constraint. This will be important in what follows.

We next assume the existence of a Birkhoff theorem, so that in vacuum the solution can be put in Schwarzschild form (\ref{eq:schwarz metric}) with $M=constant$. In this case, there must be a linear combination of the gravitational Hamiltonian and diffeomorphism constraints such that:
\be
A H^{(G)}  + BH^{(G)}_r = -M',
\label{eq:linear comb}
\ee
so that on the constraint surface $M'=0$. The coefficients depend on the phase space parameters but must be non-degenerate almost everywhere on the constraint surface. By virtue of (\ref{eq:FRM}) and (\ref{eq:GHam}), respectively, $M'$ and the coefficients $A,B$ are independent of $P_R$. They do, however, depend on $(\Lambda\, ,P_\Lambda)$ and hence on $(M\, ,P_M)$.

Using (\ref{eq:linear comb}) the full Hamiltonian in the presence of matter can  be written:
\bea
H_{TOT} &=& \frac{N}{A}\left(-M' - B H^{(G)}_r + A \frac{\rho^{(m)}}{\Lambda}\right) + N_r (\tilde{P}_RR' + P_M M' + \Pi_\psi \psi ') \none
 &=& \frac{N}{A}\left(- M' + B \Pi_\psi \psi'  + A \frac{\rho^{(m)}}{\Lambda}\right) +
  \left(N_r -\frac{NB}{A} \right)(\tilde{P}_RR' + P_M M' + \Pi_\psi \psi ').
  \label{eq:total ham}
\eea
One finally gets the following very general and intuitive result for the Hamiltonian constraint:
\be
\tilde{H}= -M'+ \frac{A}{\Lambda}\rho^{(m)} + B \Pi_\psi \psi'\approx 0.
\label{eq:Mprime}
\ee
It is intuitive because it says that on the constraint surface the gradient of the mass function equals the energy density of the matter fields ($H^{(m)}$) plus a gravitational self interaction term \cite{DGK}.
The above is a more general version of the result in \cite{DGK} for generic 2D dilaton gravity, which includes Einstein gravity with cosmological constant in any dimension. 

\section{Gauge Fixing and the Local Hamiltonian}
Without loss of generality, choose the spatial coordinate to depend only on the areal radius, so that $R=R(r)$. The consistency condition $\dot{R}=0$ then determines the shift via  the total Hamiltonian (\ref{eq:total ham}):
\be
N_r=N\frac{B}{A}.
\label{eq:shift}
\ee
The partially reduced Lagrangian is then:
\be
L_{pr} = \int dr \left(P_M\dot{M} +\Pi_\psi \dot{\psi} - \frac{N}{A}\tilde{H}\right),
\label{eq:Lp}
\ee
where $\tilde{H}$ in (\ref{eq:Mprime}) depends only on $(\psi,\Pi_\psi,M,P_M)$.

Given the above generic geometrodynamic structure we can now complete the gauge fixing by choosing the mass function
as follows:
\be
\chi = M-f(r,t)\approx 0.
\label{eq:gauge choice}
\ee
for some suitable function $f(r,t)$. Note that a more general choice, of the form:
\be
M(r,t) = g(\psi) f(r,t),
\label{eq:more general}
\ee
for arbitrary $g(\psi)$, as done in \cite{AGP}, is also possible. (\ref{eq:more general}) can be implemented by performing the following canonical transformation in (\ref{eq:Lp}):
\bea
\tilde{M}&=&M/g(\psi); \qquad P_{\tilde{M}}= {g(\psi)}P_M;\none
\psi &=& \psi; \qquad \tilde{\Pi}_\psi = \Pi_\psi + P_{\tilde{M}} \tilde{M}\frac{g'}{g},
\label{eq:tilde M}
\eea
and then making the gauge choice $\tilde{M}=f(r,t)$. The only change is that $\tilde{H}$ in (\ref{eq:Lp}) becomes a more complicated function of $P_{\tilde{M}}$ which in turn makes the algebra messier. We henceforth restrict consideration to the simpler case (\ref{eq:gauge choice}).

It is important to note that (\ref{eq:gauge choice}) does not provide a good gauge fixing condition in vacuum since $M$ commutes weakly with the gravitational part of the Hamiltonian constraint. However when the matter fields are not zero, $M$ does not have zero Poisson bracket with the Hamiltonian constraint by virtue of the latter's dependence on $P_M$ through the coefficients $A$ and $B$. In this case, one can use the consistency condition $\dot{\chi}\approx 0$ to determine the lapse and the  Hamiltonian constraint $\tilde{H}\approx 0$ determines $P_M$ in terms of the matter phase space variables $\psi, \Pi_\psi$ and the (gauge fixed) mass function $M$.
The fully reduced Lagrangian density is therefore:
\bea
L_{red} 
  &=& \Pi_\psi\dot{\psi} + P_M(\psi,\Pi_\psi)\dot{M},
\eea
with corresponding reduced Hamiltonian density:
\be
H_{red}=-P_M(\psi,\Pi_\psi)\dot{M}
\label{eq:true1}
\ee
By inspection of $\tilde{H}$ it is clear that the reduced Hamiltonian will be a local, algebraic, time dependent function of the matter fields. Here and in the following we use $M(r,t)$ to denote the gauge fixed mass function, so that it is no longer a phase space variable.

In order to make the above construction explicit and verify its broad range of applicability, we now turn to generic Lovelock gravity in D dimensions.

\section{$P$-th Order Lovelock Gravity in D Dimensions}
The action for generic $p$-th order Lovelock gravity in D dimensions can be written\cite{lovelock}:
\bea
I&=&\frac{1}{2\kappa_n^2}\int d^nx\sqrt{-g}\sum^{[n/2]}_{p=0}\alpha_{(p)} {\cal L}_{(p)},
\label{eq:lovelock action}\\
{\cal L}_{(p)}&:=& \frac{p!}{2^p}\delta^{\mu_1...\mu_p\nu_1..\nu_p}_{\rho_1...\rho_p\sigma_1..\sigma_p}
    R_{\mu_1\nu_1}{}^{\rho_1\sigma_1}...R_{\mu_p\nu_p}{}^{\rho_p\sigma_p},
 \label{eq:lovelock lagrangian}
\eea
where
$
\delta^{\mu_1...\mu_p}_{\rho_1...\rho_p}
:=
\delta^{\mu_1}_{[p_1}...\delta^{\mu_p}_{p_p]}\,
$.
Each $\alpha_{(p)}$ is a coupling constant of dimension (length) ${}^{2(p-1)}$, with $\alpha_{(0)}$ proportional to the cosmological constant. The $p=1$ term gives the Einstein-Hilbert contribution.
The specific combination of curvature terms in each of the ${\cal L}_{(p)}$ guarantees that the equations for Lovelock gravity are second order in the metric components and ghost-free. As mentioned above the generic theory obeys a generalized Birkhoff theorem~\cite{gen birkhoff,HM11}

For Lovelock gravity, it has been shown\cite{MKT} that (\ref{eq:linear comb}) is satisfied with
\be
A= \frac{R'}{\Lambda} ; \qquad B=\frac{-y}{\Lambda}=\frac{P_MF}{\Lambda^2},
\label{eq:AB}
\ee
where
\be
\Lambda^2 = \frac{1}{F}(R'^2 - F^2 P_M^2).
\label{eq:Lambda2}
\ee
as implied by (\ref{eq:FRM}) and (\ref{eq:PM2}).
After setting $R=r$, the consistency condition $\dot{R}=0$ determines the shift to be:
\be
N_r= P_MF\frac{N}{\Lambda}.
\label{eq:shift2}
\ee
The partially reduced action is given in terms of geometrodynamic variables by:
\be
L_p(t) =\int dr \left\{ P_M\dot{M} +\Pi_\psi \dot{\psi} - \frac{NR'}{\Lambda}\left(-M'\Lambda^2+ \rho^{(m)} +  P_M F \Pi_\psi \psi'
\right)\right\}.
\label{eq:Lp2}
\ee
For completeness we write down the equations of motion for the scalar field:
\bea
\dot{\psi} &=& -\frac{N}{\Lambda} \left(\frac{\Pi_\psi}{R^{D-2}}
   + P_M F \psi'\right);\none
\dot{\Pi}_\psi &=& -\left(\frac{N}{\Lambda} \left[R^{D-2}\psi' +P_MF\Pi_\psi\right]\right)^{'}.
\label{eq:eom}
\eea
At this stage one can choose a variety of different gauges: $P_M =0$ yields Schwarzschild-type coordinates, while $\Lambda=1$ yields flat slice or Painleve-Gullstrand coordinates\cite{PG}. In order to obtain a local Hamiltonian we now specify the mass function as in (\ref{eq:gauge choice}). The consistency condition $\dot{\chi}\approx0$ for this gauge choice determines the lapse function:
\be
\frac{N}{\Lambda}= \frac{\dot{M}}{ 2 M' P_M F +F \Pi_\psi \psi'}.
\label{eq:lapse}
\ee
The gauge condition is only valid if $\{\chi,\tilde{H}\}\neq 0$, which, for $\Lambda^2\neq 0$ reads:
\be
2 M' P_M F +F \Pi_\psi \psi' \neq 0.
\label{eq:bad gauge}
\ee
As long as the above inequality holds, one can set the gauge fixing condition and the hamiltonian constraint strongly to zero. The Dirac brackets in this case equal the original Poisson brackets since both $\psi$ and $\Pi_\psi$ commute with the gauge fixing condition in this parametrization.
This yields the true Hamiltonian in (\ref{eq:true1}) where $P_M$ is a local function of $(\psi, \Pi_\psi)$ obtained by solving the Hamiltonian constraint:
\be
\frac{M'}{F}(1 - F^2 P_M^2) =  \rho^{(m)}+P_MF{\Pi_\psi\psi'}.
\label{eq:final ham}
\ee
The solution to this quadratic equation for $P_M F$ is:
\be
P_M F = \frac{-\Pi_\psi\psi'\pm\sqrt{(\Pi_\psi\psi')^2-\frac{4M'}{F}(\rho^{(m)}-M'/F)}}{2M'/F}.
\label{eq:PM3}
\ee
For the classical theory, we consider only the Hamiltonian density with the positive root of the discriminant.  Indeed, as we will see in the next Section this gives the result that in the case when $\rho^{(m)}-M'/F$ is small, one can define a perturbative regime in which, up to an additive constant, the Hamiltonian density $\sim \rho^{(m)}$, .

The consistency condition (\ref{eq:bad gauge}) implies that discriminant $\Delta$ under the square root in (\ref{eq:PM3}) not vanish. Given that we want the geometrical variables to be real, we require the stronger condition:
\bea
0<\Delta&:=&(\Pi_\psi\psi')^2-\frac{4M'}{F}(\rho^{(m)}-M'/F)\none
&=&\left(\frac{1}{4}\left(\frac{\Pi^2_\psi}{R^{D-2}}+R^{D-2}\psi'^2\right)-\frac{2M'}{F}\right)^2-\left[\frac{1}{4}\left(\frac{\Pi^2_\psi}{R^{D-2}}-R^{D-2}\psi'^2\right)^2\right]
\label{eq:Delta}
\eea
The above is of the form
\be
|A^2+B^2-2M'/F|\geq |A^2-B^2|,
\ee
with $A^2=\frac{\Pi^2_\psi}{2R^{D-2}}$ and $B^2= R^{D-2}\psi'^2/2$, so that $\Pi_\psi \psi' = 2AB$.
Clearly a sufficient condition for this is that $2M'/F<0$. More generally, for a given set of initial data, at $t=0$, say, it is relatively straightforward to implement a valid gauge condition by making a judicious choice of $M'/F|_{t=0}=P(r)$\footnote{Note that $F=1-2M/r$, so this yields a linear ODE for $M$ that has a solution as long as $P(r)/r$ is integrable on the half-line $[r_0,\infty)$.}. One possibility, which will be relevant for what follows, is simply:
\be
\left.\frac{M'}{F}\right|_{t=0} = \rho^{(m)}+\epsilon(r)/2= A^2+B^2+\epsilon(r)/2,
\label{eq:Mchoice}
\ee
where $\epsilon(r)$ is everywhere positive.
The consistency then reduces to:
\be
|A^2+B^2+\epsilon|- |A^2-B^2|>0,
\ee
which is identically satisfied. We therefore claim that it is in principle possible to find a suitable gauge fixing condition of the form (\ref{eq:gauge choice}) for generic initial data. The term ``suitable'' refers to the validity of the gauge choice (\ref{eq:bad gauge}) and the positivity of the square root at time $t=0$. This does not address the issue of boundary conditions nor the possibility that the gauge may break down after some finite time $t$. Boundary conditions will be discussed in the next Section.

The resulting reduced action, given explicitly in terms of $(\psi,\Pi_\psi)$, is remarkably simple and universal in form:
\bea
L(t)&=& \int dr \left(\Pi_\psi\dot{\psi}- \frac{\dot{M}}{M'} \frac{M'}{F} (P_M F)\right) \none
  &=& \int dr \left(\Pi_\psi\dot{\psi}- \frac{\dot{M}}{2M'}
   \left[-\Pi_\psi\psi'+\sqrt{(\Pi_\psi\psi')^2-\frac{4M'}{F}(\rho^{(m)}-M'/F)}\right]\right).
   \label{eq:reduced L}
\eea
This is the main result of our paper.

We note that knowledge of the scalar field and its conjugate on a given slice allows the metric functions $N,N_r$ and $\Lambda^2$ in terms of the scalar field and its conjugate to be determined using (\ref{eq:lapse}), (\ref{eq:shift2}) and (\ref{eq:Lambda2}), respectively:
Thus the complete spacetime can in principle be reconstructed from the data $(\psi,\Pi_\psi)$ on each spatial slice, as long as none of the expressions diverge or become imaginary. Since, as argued above, the gauge choice (\ref{eq:gauge choice}) is invalid in vacuum, i.e. when the matter field and its conjugate vanish, we now turn to the important issue of boundary conditions.

\section{Boundary Conditions}

For the matter field, the usual boundary conditions consistent with asymptotic flatness in $D$ space-time dimensions are usually chosen to be:
\bea
\psi\to Br^{-\frac{D-3}{2}-\frac{\epsilon}{2}};\none
\Pi_\psi \to C r^{\frac{D-3}{2} -\frac{\epsilon}{2}},
\label{eq: matter falloff}
\eea
so that
\bea
\rho^{(m)} \sim \Pi_\psi\psi' \to r^{-1-\epsilon};\none
\Pi_\psi \dot{\psi}\to C\dot{B}{ r^{-\epsilon}}.
\eea
The only boundary terms that arise in the variation of (\ref{eq:reduced L}) come from the variation of $\psi'$. For $\epsilon>1$, these will be finite as long as $\Delta\neq0$, as required by the consistency condition on the gauge choice, and $\dot{M}/M'$ is regular at the boundary.

Since the gauge fixing condition is problematic in vacuum, one might expect corresponding subtleties to arise in asymptotically flat space times. These subtleties are revealed by  inspection of of the metric functions. In fact (\ref{eq:final ham}) implies that  if $\rho^{(m)}\to0$ faster than $M'$, then $FP_M$ must go to unity. This in turn yields $\Lambda^2\to 0$, which signals a breakdown in the coordinates. In addition $M'\to 0$ faster than the matter fields is ruled out by (\ref{eq:final ham}) since it would required $|1-F^2P_M^2|>1$, which would result in $\Lambda^2<0$. The question then arises: for what physically relevant calculations is the local Hamiltonian useful?

One possibility is to avoid asymptotically flat spacetimes and explore more general situations, such as those containing ingoing/outgoing radiation. Alternatively, one can note that the perturbative regime (for which the scalar obeys the usual free field equations) corresponds to:
\be
\rho^{(m)}-M'/F \to 0,\label{eq:fo1}
\ee
as follows from (\ref{eq:reduced L}).
In this case, for the $+$ sign in (\ref{eq:reduced L}), we find
\be
-\frac{\dot{M}}{M'} \frac{M'}{F} (P_M F)\to \frac{\dot{M}}{F}
   \frac{\rho_m-M'/F}{\Pi_\psi\psi'}\to 0.
\ee
Since $P_M\to0$ this is consistent with the expression for $N/\Lambda$ above, i.e. in this limit
\be
\frac{N}{\Lambda} \to \frac{\dot M}{F\Pi_\psi \psi'}.
\ee
Moreover, in this limit $N_r\to0$ and $\Lambda^2 \to 1/F$. So if we can choose
\be
\frac{\dot{M}}{F} \to \Pi_\psi \psi',\label{eq:fo2}
\ee
everything appears consistent.

This suggests the following procedure for fixing initial data in calculations of spherical self-gravitating collapse:
First assume that the matter obeys the fall off conditions (\ref{eq: matter falloff}) at $t=0$. Then choose gauge fixing to be of the form:
\be
M = a(r) + b(r)t.
\ee
Finally determine the functions $a(r)$ and $b(r)$ so that on the initial slice:
\bea
\left.\frac{M'}{F}\right|_{t=0}&=& \frac{a'(r)}{1-a(r)/r}=\rho^{(m)} + O(r^{-\alpha});\none
\left.\frac{\dot{M}}{F}\right|_{t=o} &=& \frac{b(r)}{1-a(r)/r}=\Pi_\psi \psi'+ O(r^{-\alpha}).
\eea
Note that this choice of $M(r,0)$ matches (\ref{eq:Mchoice}) in the previous section. One can then evolve the matter field according to the equations of motion (\ref{eq:eom}) until the gauge fixing condition breaks down or one of the metric functions diverge. Note that from the equations of motion for the matter field, the asymptotic time evolution for the matter field is governed by:
\bea
\dot{\psi}\to r^{-3/2-\epsilon};\none
\dot{\Pi}_\psi \to r^{-3/2-\epsilon}.
\eea
This suggest thats the asympotic relationship between matter and gauge fixing will be preserved for a finite time. 


\section{Conclusion}
We have shown that a local Hamiltonian can be obtained for spherically symmetric massless scalar field dynamics in any theory of gravity to which the geometrodynamics of Kuchar can be applied. The class of theories for which this holds includes the generic Lovelock case as well as generic 2D dilaton gravity.
The reduced action (\ref{eq:reduced L}) is remarkably simple and universal for all Lovelock theories in any dimension. The existence of a local Hamiltonian is a direct consequence of Eq.(\ref{eq:GHam}) which shows that the gravitational Hamiltonian is linear in $P_R$. In fact there are two terms, one each of  $H^{(G)}$ and $H^{(G)}_r$. The coefficients $A$ and $B$ that determine (\ref{eq:final ham}) are, up to an overall normalization factor, precisely those required to cancel the two linear terms: $A R'+By=0$. The overall normalization factor should be the same for all such theories, and is such as to yield $N\Lambda/R'$ as the coefficient of $M'$ in
(\ref{eq:total ham}). This coefficient plays a role in determining the boundary conditions, since the boundary term in the Hamiltonian comes from the variation of $M'$.
We conjecture that the form of the true Hamiltonian (\ref{eq:PM3}) in this type of gauge is the same for all theories that obey a Birkhoff theorem and possess a mass function $M$ that is constant on the constraint surface in vacuum. The gravitational back reaction is determined completely by the dependence of $F$ on $M$, as one may expect from physical grounds. The extension to theories with cosmological constant is straightforward, since it merely modifies the relationship between $F$ and $M$.

We note that our geometrodynamic formulation and family of gauge choices results in a reduced Hamiltonian that is much simpler than that of [10].  This makes it easier to discuss both the advantages and potential disadvantages of this family of gauges. Specifically, we were able to show in Section IV that the reality  condition on the square root is directly tied to the validity of the gauge choice and more importantly, we were able to find consistent gauge choices for generic initial data.

Despite the universality and elegance of the final reduced Hamiltonian, there are potential problems in implementing this gauge fixing procedure in asymptotically flat spacetimes, related to the fact that the gauge condition is not valid in vacuum. We have proposed one possible approach to this problem for the investigation of spherical collapse, but the viability of this approach requires further investigation. As mentioned in Section III, another possibility is to consider more complicated gauge fixing conditions that depend on both $\psi$ and $M$, as done in [10].

Finally, we note another potential application for the local Hamiltonian, namely the calculation of scattering amplitudes for spherically symmetric self-gravitating matter using the phase space path integral. In principle, the relatively simple algebraic form of the local Hamiltonian might provide simplifications over the method used in \cite{fisher}, for example, to calculate the tree level four point interaction. However, it should be noted that in the case of Hamiltonians quadratic in the momenta it is possible to perform the integral over the momentum to obtain a simple covariant path integral with a Lagrangian that is quadratic in the velocities. In the present case, the integration over the momentum is not trivial and given the non-polynomial nature of the Hamiltonian, it will yield an expression that is a non-polynomial function of the velocity. It is nonetheless an intriguing question that deserves further attention.

\section{Acknowledgments} This work was supported in part by the Natural Sciences and Engineering Research Council of Canada. We thank V. Husain , J. Pullin  and W. Unruh  for useful comments. G.K. is greatful for invaluable discussions and the collaboration of Hideki Maeda and Tim Taves on the results concerning Lovelock gravity that have been used in the present paper.


\begin{thebibliography}{99} 

\bibitem{string}For a recent review, see S. Wadia,``String Theory: A Framework for Quantum Gravity and Various Applications''	TWAS Jubilee Publication, [ArXiv:hep-th/0809.1036v2].
\bibitem{lqg} For a recent review, see M. Bojowald,
``Loop Quantum Gravity and Cosmology: A dynamical introduction''
based on a talk presented at the workshop "Foundations of Space and Time - Reflections on Quantum Gravity", STIAS, Stellenbosch, South Africa, 10-14 August 2009, to appear in CUP,
[ArXiv:gr-qc/1101.5592].
\bibitem{time}  For an interesting current attempt, and for references to the earlier literature see V. Husain and T. Pawlowski, ``Time and a physical Hamiltonian for quantum gravity'', [ArXiv:gr-qc/1108.1145].
\bibitem{berger} B. Berger, D. Chitre, Y. Nusku, V. Moncrief, Phys. Rev. {\bf D5} 2467 (1972).
\bibitem{unruh} W.G. Unruh, Phys. Rev. {\bf D14} 870 (1976).
\bibitem{choptuik} M.W. Choptuik, Phys. Rev. Lett. {\bf 70} (1) 9 (1993).
\bibitem{vacuum} J. Gegenberg and G. Kunstatter, Phys. Rev. {\bf D47} 4192 (1993);   H.A. Kastrup and T. Thiemann, Nucl. Phys. {\bf B399} 211 (1993); {\bf B425} 665 (1994)
\bibitem{viqar2} V. Husain and O. Winkler, Phys.Rev. {\bf D71}  104001 (2005) [arXiv:gr-qc/0503031]
\bibitem{DGK} R. Daghigh, J. Gegenberg and G. Kunstatter, Class. Quant. Grav. {\bf 24}:2099 (2007) [arXiv:gr-qc/0607122]
\bibitem{AGP} N. Alvarez, R. Gambini and J. Pullin, 
Phys. Rev. Lett. {\bf 108}  051301 (2012)
[arXiv:1111.4962v2 [gr-qc]].
\bibitem{BS}M. Bojowald and R. Swiderski, Class. Qu. Grav. {\bf 24}, 2129 (2006) [arXiv:gr-qc/0511108]
\bibitem{viqar} V. Husain, private communication.
\bibitem{bill} W.G. Unruh, private communication.
\bibitem{grumiller} For a review see D. Grumiller, W. Kummer and D.V. Vassilevich Phys.Rept.{\bf 369} 327,(2002) [arXiv:hep-th/0204253]
\bibitem{lovelock} D. Lovelock, ``The Einstein tensor and its generalizations'', J. Math. Phys. 12 (1971) 498.
\bibitem{hideki} H. Maeda and M. Nozawa, PHys. Rev. {\bf D77}:064031 (2008) [arXiv:0709.1199]; H. Maeda, S. Willison and S. Ray, Class. Quant. Grav. {\bf 28}:165005 (2011) [arXiv:1103.4184]
 \bibitem{string lovelock}B. Zwiebach, Phys. Lett. B 156, 315 (1985). See also D. G. Boulware and S. Deser,
Phys. Rev. Lett. 55, 2656 (1985) and B. Zumino, Phys. Rept. 137, 109 (1986).
\bibitem{rizzo}T.G. Rizzo, Collider Production of TeV Scale Black Holes and Higher-Curvature Gravity``, JHEP0506:079(2005) [ArXiv:hep-ph/0503163].
\bibitem{gen birkhoff}
R.~Zegers, J. Math. Phys. {\bf 46}, 072502 (2005). See also S. Deser and J. Franklin, Class.Quant.Grav.22:L103-L106 (2005)
\bibitem{HM11} H. Maeda, S. Wilson and S. Ray, Class. Quant. Grav. {\bf 28}, 165005 (2011).
\bibitem{misner_sharpe} C.W. Misner and D.H. Sharp, Phys. Rev. {\bf 136}, B572 (1964).
\bibitem{kuchar} K. Kuchar, Phys. Rev. {\bf D50} (6) 3961 (1994) [arXiv:gr-qc/9403003]
\bibitem{GKM} The mass function and the identification of its conjugate momentum was obtained for generic 2-D dilaton gravity by J. Gegenberg, G. Kunstatter and D. Louis-Martinez, Phys. Lett. {\bf B321} 193 (1994) [arXiv:gr-qc/9309018]
\bibitem{LSW} J. Louko, J.Z. Simon and S.N. Winters-Hilt, Phys. Rev. {\bf D55} (6) 3525 (1997).
\bibitem{TLKM} T. Taves, D. Leonardo, G. Kunstatter and R.B. Mann, ``Hamiltonian formulation of scalar field collapse in Einstein-Gauss-Bonnet Gravity'', Class. Quant. Grav. {bf 29} 015012 (2011) [arXiv:1110.1154].
\bibitem{MKT}  G. Kunstatter, H. Maeda and T. Taves, ``Geometrodynamics of generic Lovelock gravity'', [arXiv:1201.4904]
\bibitem{PG} P. Painlev\'e, ``La m\'ecanique classique et la th\'eorie de la relativit\'e", C. R. Acad. Sci. (Paris) {\bf 173} 677 (1921); A. Gullstrand, ``Allgemeine L\"osung des statischen Einkörperproblems in der Einsteinschen Gravitationstheorie", Arkiv. Mat. Astron. Fys. {\bf 16}(8) 1 (1922).
\bibitem{fisher} P. Fisher, D. Grumiller, W. Kummer and D.V. Vassilevich,  Phys.Lett. B521 (2001) 357-363, Erratum-ibid. B532 (2002) 373 [arXiv:gr-qc/0105034v2]
 \end{thebibliography}
\end{document}